\documentclass[10pt,journal,compsoc]{IEEEtran}
\usepackage {amsmath}
\usepackage{footmisc}
\usepackage {amssymb}
\usepackage{multirow}
\usepackage{fixltx2e}
\usepackage[T1]{fontenc}
\usepackage{microtype}
\usepackage{booktabs}
\usepackage{tabularx}
\usepackage{color}

\usepackage{ifpdf}

\ifCLASSOPTIONcompsoc
  \usepackage[nocompress]{cite}
\else
  \usepackage{cite}
\fi

\ifCLASSINFOpdf
  \usepackage[pdftex]{graphicx}

\else

   \usepackage[dvips]{graphicx}

\fi

\usepackage{amsmath}

\usepackage{acronym}
\usepackage[ruled,linesnumbered]{algorithm2e}
\usepackage{array}
\usepackage{mdwmath}
\usepackage{mdwtab}
\usepackage{eqparbox}

\ifCLASSOPTIONcompsoc
  \usepackage[caption=false,font=footnotesize,labelfont=sf,textfont=sf]{subfig}
\else
  \usepackage[caption=false,font=footnotesize]{subfig}
\fi
\usepackage{fixltx2e}

\ifCLASSOPTIONcaptionsoff
  \usepackage[nomarkers]{endfloat}
 \let\MYoriglatexcaption\caption
 \renewcommand{\caption}[2][\relax]{\MYoriglatexcaption[#2]{#2}}
\fi

\usepackage{url}

\ifCLASSINFOpdf
  \usepackage[pdftex]{thumbpdf}
\else
  \usepackage[dvips]{thumbpdf}
\fi

\begin{document}



\title{Chiplet-Gym: Optimizing Chiplet-based AI Accelerator Design with Reinforcement Learning  }

\author{Kaniz~Mishty~\IEEEmembership{}
        and~Mehdi~Sadi,~\IEEEmembership{Member,~IEEE,}
\IEEEcompsocitemizethanks{\IEEEcompsocthanksitem The authors are with the Department of Electrical and Computer Engineering, Auburn University, Auburn, AL 36849 USA \protect\\
E-mail: kzm0114@auburn.edu; mehdi.sadi@auburn.edu
}
\thanks{This work was supported in part by the National Science Foundation (NSF) under Grant Number CRII-2153394.}
\thanks{Manuscript received February 03, 2024; revised May 30, 2024.}}

%
%

\markboth{IEEE Transactions on Computers,~Vol.~XX, No.~X, August~202X}%
{Shell \MakeLowercase{\textit{et al.}}: Bare Advanced Demo of IEEEtran.cls for IEEE Computer Society Journals}
%



\IEEEtitleabstractindextext{%
\begin{abstract}

Modern Artificial Intelligence (AI) workloads demand computing systems with large silicon area to sustain throughput and competitive performance.  However, prohibitive manufacturing costs and yield limitations at advanced tech nodes and die-size reaching the reticle limit restrain us from achieving this.  With the recent innovations in advanced packaging technologies, chiplet-based architectures have gained significant attention in the AI hardware domain. However, the vast design space of chiplet-based AI accelerator design and the absence of system and package-level co-design methodology make it difficult for the designer to find the optimum design point regarding Power, Performance, Area, and manufacturing Cost (PPAC). This paper presents Chiplet-Gym, a Reinforcement Learning (RL)-based optimization framework to explore the vast design space of chiplet-based AI accelerators, encompassing the resource allocation, placement, and packaging architecture. We analytically model the PPAC of the chiplet-based AI accelerator and integrate it into an OpenAI gym environment to evaluate the design points. We also explore non-RL-based optimization approaches and combine these two approaches to ensure the robustness of the optimizer. The optimizer-suggested design point achieves $1.52\times$ throughput, $0.27\times$ energy, and $0.01\times$ die cost while incurring only $1.62\times$ package cost of its monolithic counterpart at iso-area. 


\end{abstract}

\begin{IEEEkeywords}
AI Accelerator, Chiplet, Heterogeneous Integration, Design Space Exploration, Reinforcement Learning.
\end{IEEEkeywords}}

\maketitle

\IEEEdisplaynontitleabstractindextext

\IEEEpeerreviewmaketitle

\section{Introduction}
\label{sec1:introduction}

As Large Language Models (LLMs), such as chatGPT, GPT-4, LLaMA \cite{LLM}, etc.,
gain widespread use, there is a growing demand for energy-efficient hardware that can deliver high throughput. To support hundreds of trillions of operations and hundreds of Giga Bytes of data movement,  the high-performance and energy-efficient
hardware demands more silicon area, accommodating more
compute cores and memory capacity. Training any state-of-the-art AI or Deep Learning (DL) model with a single GPU or accelerator is nearly impossible due to extreme computing and memory demands. The data centers are equipped with clusters of powerful computers and GPUs connected via PCIe, NVLink, etc.\cite{usenix}\cite{hir}. 
Even though these supercomputers can deal with large workloads, they consume a significant amount of energy \cite{usenix} and involve longer latency. Because off-board communications consume at least one order of magnitude more power and time than any on-package communications \cite{off_board_power}. The ideal
scenario would be a hardware capable of housing the
entire model parameters and intermediate activations on-chip\cite{chiplet_cloud}, promising optimal performance and energy efficiency. Unfortunately, this is not feasible due to the stagnation of Moore's law and Dennerd's scaling, die size reaching the reticle limit, and the prohibitive manufacturing cost and yield limitations\cite{hir}. Consequently, researchers endeavor to replicate this `hypothetical ideal' hardware concept by integrating multiple smaller chiplets at the package level, allowing near-ideal performance while minimizing costs and energy consumption. 

With the advent of advanced packaging technologies, the chiplet-based heterogeneous integration has opened up a new dimension of chip design, \textbf{More-than-Moore} \cite{hir}. In chiplet-based system, multiple chiplets (i.e., SoCs) of diverse functionalities (e.g., logic dies, memories, analog IPs, accelerators etc.) and tech nodes (e.g., 7nm or beyond) from different foundries are interconnected in package level using the advanced packaging technologies, such as CoWoS, EMIB,
etc. \cite{hir}. The value proposition of chiplet-based architectures is manifold. Compared to multiple monolithic SoCs interconnected via off-package or off-board links such as PCIe, NVLink, CXL etc. \cite{hir}, package-level integration of multiple monolithic SoCs via 2.5D or 3D has accelerated performance and lower energy consumption alleviating off-package communications. Chiplet-based systems offer lower RE (Recurrent Engineering) cost by providing higher yield and lower NRE (Non-Recurrent Engineering) by enabling IP reuse and shortening IC design cycle \cite{chiplet_actuary}.

The commercial chiplet-based general purpose products \cite{intel_ponte_vecchio} \cite{AMD_epyc2} are designed and developed at vertically integrated companies without exposing much knowledge about the chiplet-based architectures' design space. Unlike these general purpose products, chiplet-based AI accelerators demand extensive design space exploration to hit the target Power, Performance, Area, and Cost (PPAC) budget. From architectural perspective, designers must consider the resource allocation, mapping and dataflow of the DNN workloads. From communication and integration perspective, chiplet placement, routing protocols, stacking/packaging technologies, interconnect types, and finally from application perspective, system requirement, such as reliability, scalibility etc., should be considered all at the same time while optimizing for PPAC \cite{monad}. The existing works often focus either on the architectural or integration aspects as a separate design flow: explore routing and packaging given chiplets \cite{NoP_chiplet}\cite{2.5D_placement}\cite{RL_NoC}
 or explore chiplets architecture given the packaging \cite{chiplet_cloud}\cite{shao2019simba}\cite{nn_baton} \cite{sprint}. 
An isolated approach, addressing individual aspects independently, may result in sub-optimal designs due to the inter-dependency among these factors. For instance, varying resource allocation impacts communication demands, influencing the choice of packaging and its configuration, consequently leading to cost variations.

Currently, many flavors of packaging technologies, both from 2.5D and 3D, are available from the industry leaders, which makes it difficult for system designers and integrators to choose the optimum set of configurations from the vast design space based on the system requirements \cite{hir}. The various packaging technologies differ in fabrication cost and complexity, performance, and underlying integration technologies \cite{hir}. As a result, no single package technology can be marked as superior to others. Each of the other domains, such as resource allocation, chiplet granularity, placement, Network on Package (NoP), and interconnect architectures, to name a few, also has an extensive design space. A proper co-optimization across all these domains based on the system and application requirements at the available cost is necessary for a successful chiplet based system design. 
Optimizing all possible domains results in a combinatorial explosion where brute force search is not an option and random search might not result in the optimum point. The expensive simulation environment of chip design exacerbates this problem.

To overcome these limitations, in this paper, we make the following contributions bridging the gap between the system requirements and design aggregation, planning, and optimization for chiplet-based architecture.

\begin{itemize}
    \item We develop a co-design methodology for chiplet-based AI accelerators. The co-design task contemplates resource allocation, such as the number of AI chiplets, memory capacity, and bandwidth; partitioning and placement of chiplets such as aspect ratio of the accelerator chiplet arrays, and logical placement of accelerator and memory  chiplet; different packaging technologies (i.e., CoWoS, EMIB, SoIC, and FOVEROS \cite{hir}) and their attributes such as bandwidth, bump pitch density, cost and complexity, to optimize the system-level Power, Performance, Area, and Cost (PPAC) of the chiplet-based AI accelerators.
    
    \item We formulate an analytical cost model for assessing the chiplet-based architectures. This analytical model enables us to assess the chiplet-based AI accelerator in a time-and-resource-constrained environment.


    \item To optimize throughput, energy efficiency, and cost, we identify the inter-dependency of the design space parameters and formulate the optimization problem as  a Reinforcement Learning (RL) problem. We also explore non-RL based optimization approaches, such as simulated annealing, and combine these two approaches to ensure the robustness of the optimizer. 

    \item Finally, we validate our methodology by comparing the performance of our optimized design against state-of-the-art monolithic GPU on MLPerf benchmark and justify the performance improvement.
\end{itemize}

The rest of the article is organized as follows. Section \ref{sec2:background_and_motivation} presents the background. Section \ref{sec3:thruput_energy_formulation} describes the analytical modeling and design space exploration. The optimization framework is presented in section \ref{sec4:optimizer_framework} followed by experiments and results in Section \ref{sec5:experiment_and_result}, related works in Section \ref{sec6:related_works}, limitations and future works in Section \ref{sec7:lim_n_future_wrks} and conclusion in Section \ref{sec8:conclusion}. 
\section{Background}
\label{sec2:background_and_motivation}
\subsection{AI workloads and Accelerators}
\subsubsection{AI workloads}
The primary domains of AI encompass Computer Vision (CV), Natural Language Processing (NLP), Recommender Systems, and Reinforcement Learning. The integration of these domains has led to the emergence of Generative AI, enabling models to generate diverse content, including text and images. In Generative or Multi-modal AI, diverse AI/DNN (Deep Neural Network) models are fused together to generate an output. For example, OpenAI's DALL-E\cite{dalle}, constructed on GPT-3 and CLIP model, leverages the principles of LLM and CV models \cite{dalle} to produce images from texts. While the architectural characteristics and parameters of LLM and CV models may differ, their fundamental components share similarities with the structure of Transformer\cite{vaswani2017attention} for NLP and ResNet\cite{resnet} for CV, respectively. The critical operations in CV models involve regular convolution, Depth-wise or Point-wise convolution, residual blocks, FC (Fully Connected) operations, whereas the scaled-dot product attention operations, and FC operations dominate in LLM. These operations can be expressed as or converted to matrix-matrix/vector multiplication (GEMM) with massive parallelism.

\begin{figure}[ht]
    \centering
    \includegraphics{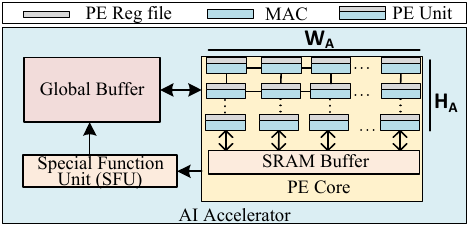}
    \caption{AI accelerator chiplet architecture} 
    \label{fig:AI_accel_arch}
\end{figure}

\begin{figure*}
    \centering
    \includegraphics[width=\textwidth]{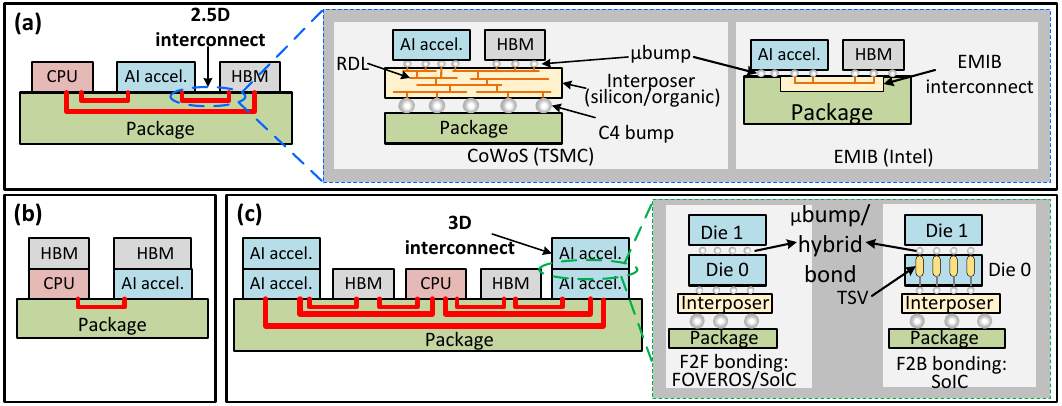}
    \caption{Top-level system architecture for different scenarios. (a) CPU, AI accelerator and HBM chiplets are connected in package level through 2.5D interconnects. CoWoS and EMIB are two options of 2.5D interconnects. (b) CPU and AI accelerator chiplets are connected through 2.5D interconnects and HBM is stacked on top of CPU and AI accelerator through 3D interconnects. (c) Two AI accelerator chiplets are stacked on top of each other through 3D interconnects and they are interconnected to CPU, HBM and other AI chiplets pair through 2.5D.} 
    \label{fig:arch_option}
\end{figure*}
\subsubsection{AI Accelerator} Systolic array \cite{tpuv4} type architecture, leveraging the inherent parallelism of DNN workloads, has been used as the core of AI accelerators. A typical AI accelerator is composed of arrays of Processing Element (PE) for computation and on-chip buffer to hold the weights and activations. PEs are composed of Multiplier-Adder (MAC) units and small register file for each MAC units to hold the stationary data, depending on the dataflow. The size of PE array, memory hierarchy, and memory size are critical design parameter of a AI accelerator. Fig. \ref{fig:AI_accel_arch} shows a AI accelerator with a PE core, Special Function Unit (SFU), and Global Buffer. The PE core contains a small SRAM buffer and bunch of PE units. Each PE consists of a MAC unit and a reg. file \cite{sot_tcad}. 

\subsection{Chiplets and Heterogeneous Integration}
\subsubsection{2.5D architecture}
\label{}
In 2.5D architecture, two or more chiplets, fabricated separately, are connected side-by-side with each other in package-level through interposer (silicon/organic) or silicon bridge. Two commercial 2.5D interconnects are Chip on Wafer on Substrate (CoWoS) from TSMC \cite{ISSCC2021forum}
and Embedded Multi-die Interconnect Bridge (EMIB) from Intel \cite{EMIB}. In CoWoS, two side-by-side dies are connected with each other and with package substrate through an intermediate interposer layer \cite{ISSCC2021forum}.
Interposer can be active and passive. Active interposer contains embedded logics and Re-Distribution Layers (RDL) where as passive interposer containing RDLs are only used for routing purpose. CoWoS typically employs passive interposer for 2.5D integration.  In contrast, Intel's EMIB utilizes thin silicon pieces with multilayer BEOL interconnects (Silicon Bridge) embedded in the organic package substrate for high-density localized interconnects, eliminating the need for a separate interposer layer \cite{EMIB}. CoWoS and EMIB architectures are illustrated in Fig. \ref{fig:arch_option} (a).

\subsubsection{3D architecture}
In 3D, two or more separately fabricated chiplets are stacked on top of each other through 3D interconnects formed with copper micro-bumps, or hybrid wafer bonding \cite{mathur2021thermal}. Depending on the bonding interface orientation of the interconnected dies, different bonding configurations are possible, such as face-to-face (F2F), face-to-back (F2B), back-to-back (B2B) etc. Intel's FOVEROS \cite{FOVEROS} uses F2F bonding where the face of the top die is bonded to the face of the bottom die (active interposer) through Cu micro-bump connections.
Bottom die is connected to the package through TSV \cite{FOVEROS}. TSMC has the option of both F2F and F2B bonding configuration in their System on Integrated Chips (SoIC), however, they use hybrid bonding instead of Cu $\mu$-bumps \cite{SoIC}. The latest upgrade of FOVEROS, FOVEROS-Direct, also leverages direct cu-cu hybrid bonding for inter-die interconnection. Recently, both 2.5D and 3D can be integrated on the same package and these architectures are known as 5.5D \cite{5p5D_testing}.
 

\section{Throughput formulation and Design space exploration}
\label{sec3:thruput_energy_formulation}
In this section, we formulate the cost model for chiplet-based AI accelerators, including throughput, energy, and cost. We perform design space exploration to comprehend the influence of various design parameters on the formulated cost model.

\subsection{Top level Architectural exploration}
We explore two architectural approaches: (i) 2.5D architecture, where all chiplets are connected with each other at the package level through 2.5D interconnects (\textcolor{black}{Fig. \ref{fig:arch_option}(a)}). (ii) 5.5D (combining 2.5D and 3D)\cite{5p5D_testing} where two or more 3D-stacked (connected via 3D interconnects) chiplets are further linked through 2.5D interconnects \textcolor{black}{(Fig. {\ref{fig:arch_option}  (b) \& (c)})}. In all cases, the architecture of the AI accelerator chiplet is a regular systolic-array composed of PE array and dedicated on-chip buffer shown in Fig. \ref{fig:AI_accel_arch}\cite{sot_tcad}. However, the number of PE units and on-chip buffer size varies with the number of allocated chiplets, as we consider a fixed package size.

\subsubsection{2.5D architecture}
In 2.5D architecture, we consider that CPU, AI accelerator, and HBM chiplets are connected at the package level through 2.5D interconnects (Fig. \ref{fig:arch_option} (a)). We explore two 2.5D integration technologies, EMIB and CoWoS, and their different configurations.

\subsubsection{5.5D architecture (combining 2.5D and 3D)} 5.5D architecture is divided into two cases: (i) memory-on-logic, where HBMs are stacked on top of CPU and/or AI chiplets as shown in Fig. \ref{fig:arch_option} (b), and (ii) logic-on-logic, where two AI chiplets are 3D-stacked on top of each other. These 3D-stacked AI chiplets are connected to CPU and/or HBM and other 3D-stacked AI chiplets through 2.5D interconnects as shown in Fig. \ref{fig:arch_option} (c). \cite{3d_gtech} explored more than 2-tiers in 3D stacks, however, we limit our exploration to only 2-tiers to avoid temperature-induced breakdowns \cite{mathur2021thermal}. We explore the off-the-shelf 3D integration techniques, SoIC and FOVEROS, and their different configurations. Depending on the integration technology and their configuration settings, these architectures offer different bandwidths, energy efficiency, area efficiency, and cost.



\subsection{Throughput and Energy efficiency formulation}
\label{}
\subsubsection{Throughput}
We define system throughput as \emph{tasks completed per second},
\begin{equation}
    \begin{aligned}
        T = \frac{tasks}{sec} 
    \end{aligned}
\end{equation}
$tasks$ represents different entities depending on the DNN domain and its mode of operations. During inference, $tasks$ represents the number of $inferences$. During training of CV and  NLP models, $tasks$ means the number of $images$ and $tokens$ processed per second, respectively. $tasks/sec$ can be decomposed into \cite{evaluate_dnn_processors}
\textcolor{black}{
\begin{equation}
    \begin{aligned}
        \frac{tasks}{sec} = \frac{ops}{sec} \times \frac{1}{(\frac{ops}{task})_{G}} \times \frac{1}{(\frac{ops}{task})_{nG}}\times M_{eff}
    \end{aligned}
\end{equation}}

Here, $ops/sec$ depends on both DNN hardware and DNN models. GEMM operations per task, $(ops/task)_{G}$, and non-GEMM operations per task $(ops/task)_{nG}$ depend on only DNN models, and $M_{eff}$, mapping efficiecny depends on DNN models and hardware, along with mapping strategies. $ops$ means the MAC operation. \textcolor{black}{The GEMM operations are performed in the systolic array. The non-GEMM operations such as softmax is performed in the SFU of the accelerator. Dropout and residual operations, manifested as Element-wise multiplication and addition, is also performed using the MAC modules. Layer normalization and other reduction or control flow operations are taken care in the ALU or scalar unit of the SFU.}

For a system comprising multiple AI accelerator chiplets, the $operations/sec$ is expressed as,
\begin{equation}
\label{eq:ops_sys}
    \begin{aligned}
        (\frac{ops}{sec})_{sys} = (\frac{ops}{sec})_{AI\_chip}\times AI\_chip_{tot}\times U_{sys}
    \end{aligned}
\end{equation}
Where $(ops/sec)_{AI\_chip}$ is the peak throughput per AI chiplet, $AI\_chip_{tot}$ = total number of AI chiplets, and $U_{sys}$ = system utilization factor. It represents the effective fraction of the active chiplets out of the total chiplets. It depends on the interchiplet communication bandwidth ($BW_{AI-AI}$), determined by choice of the packaging architecture, package type, and their different configuration. In section \ref{sec:inter_chiplet_comm_bw}, we describe this in detail. The peak throughput per AI chiplet is expressed as
\begin{equation}
    \begin{aligned}
    \label{ops_per_chip}
        (\frac{ops}{sec})_{AI\_chip} = (\frac{1}{\frac{cycles}{op}}\times \frac{cycles}{sec}) \times  PE_{tot} \times U_{AI\_chip}
    \end{aligned}
\end{equation}
Where, 
\begin{equation}
    \begin{aligned}
        \label{cyc_per_op}
        \frac{cycles}{op} = cycle_{comm} + cycle_{op^*}
    \end{aligned}
\end{equation}
$cycle_{comm}$ = chiplet-to-chiplet communication latency, $cycle_{op^*}$ = arithmetic operation latency of the chiplet microachitecture, and $cycles/sec$=$f$, frequency of the AI accelerator chiplets. $cycle_{comm}$ depends on the distance between the data source and destination. It is impacted by the chiplet allocation, chiplet array dimension (i.e., number of AI chiplets in X and Y dimension) and the physical location of the AI and HBM chiplets. $cycle_{op^*}$ depends on the micro-architecture of the chiplet (design of PE array, MAC unit) and the type of operations. We assume that all AI chiplets can operate at the same frequency and have the same architectural and functional configuration. However, the frequency of each chiplet can be further controlled based on the data frequency and location of chiplets to optimize the system throughput and energy. $PE_{tot}$ = total number of PEs per AI chiplet, and $U_{AI\_chip}$ = chiplet utilization representing the fraction of PEs utilized during computation. $U_{AI\_chip}$ depends on mapping of the AI model tasks to the accelerator.

\subsubsection{Energy efficiency}
Energy efficiency is paramount when processing DNN at edge devices and cloud data centers. Edge devices are usually constrained by battery life and thermal budget, and data centers are typically constrained by electricity bills, thermal budget, and environmental impact \cite{usenix}. Data centers are mainly focused on achieving higher throughput, which requires higher energy budget. In this work, we closely monitor energy efficiency while maximizing the throughput. 

We define energy efficiency ($E_{eff}$) of a system as \emph{tasks completed per joule}:
\begin{equation}
    \begin{aligned}
        \label{tsk_per_jl}
        E_{eff} &= \frac{tasks}{joule}
               & = \frac{1}{\frac{joules}{ops}} * \frac{1}{\frac{ops}{task}}
    \end{aligned}
\end{equation}
$joules/operations$ depends on both DNN hardware and DNN models, whereas $operations/task$ depends on only DNN \textcolor{black}{model}. We break down the energy per operations, $E_{op}$, (i.e. $joules/operations$) into its constituent parts:
\begin{equation}
    \begin{aligned}
        \label{enrgy_per_op}
        E_{op} = E_{comm} + E_{op^*}
    \end{aligned}
\end{equation}
$E_{comm}$ is the energy required to transfer data from chiplet-to-chiplet and $E_{op^*}$ is the energy to perform an arithmetic operation. $E_{comm}$ depends on the choice of packaging architectures (e.g., 2.5D, 3D) and interconnect types (e.g., EMIB, CoWoS, FOVEROS, SoIC) and $E_{op^*}$ depends on the microarchitecture.

\subsection{Chiplet allocation and Placement}
\label{subsec:chiplet_alloc}
In the context of chiplet-based accelerator design, determining the number of chiplets, area allocated to each chiplet, and their placement becomes pivotal, as they impact the throughput, energy, and cost. Here we will delve into the relationship between yield, area, cost, communication latency across various chiplet configurations.
\begin{figure}[h]
    \centering
    \includegraphics[width=0.48\textwidth]{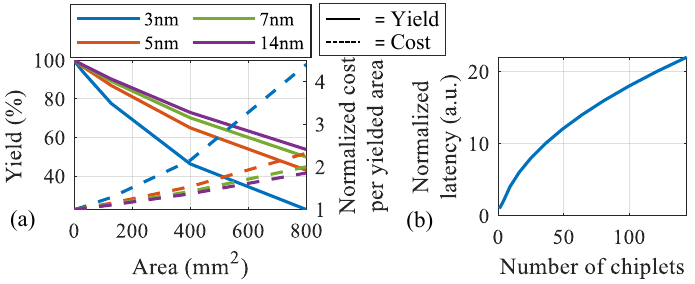}
    \caption{(a) Yield (left y-axis) and normalized cost per yielded area (right y-axis) vs area at different tech nodes. (b) Normalized latency vs number of chiplets.}
    \label{fig1:chiplet_alloc_exploration}
\end{figure}

\subsubsection{Yield and Cost vs Area} 
\label{subsubsec:yield_cost_vs_area}
Intuitively, as the chip area increases, its compute and memory capacity increases, ensuring high performance and energy-efficiency. However, as shown in Fig. \ref{fig1:chiplet_alloc_exploration} (a), we are limited by the fact that in advanced tech nodes, as the chip area increases, yield decreases, resulting in increased cost per area \cite{chiplet_actuary}. The yield of the manufactured chip, $Y_{die}$ is expressed as the following Negative Binomial model:
\begin{equation}
    Y_{chip} = (1+ \frac{dA}{\alpha})^{-\alpha}
    \label{eq1:yield}
\end{equation}
where $d$ is the defect density of the tech node, $A$ is the area of the chip, and $\alpha$ is the cluster parameter. Assuming $P_{0}$ as unit price, we can also estimate the cost per yielded area as 
\begin{equation}
    C_{yield} = \frac{P_{0}}{Y_{chip}} \approx P_{0}(1 + dA + \frac{\alpha - 1}{2\alpha}d^{2}A^{2})
    \label{eq2:cost}
\end{equation}


\subsubsection{Inter-Chiplet Communication Latency} 
As mentioned earlier, the chiplet-to-chiplet data communication latency, $cycle_{comm}$, impacts the system performance by contributing to $cycles/operations$. 
Data transfer between chiplets occurs through the package-level interconnects such as CoWoS, EMIB, FOVEROS, SoIC etc. Considering that the data might be supplied from another AI-chiplet or directly from HBM, we estimate both AI-AI chiplet communication latency, $L_{AI-AI}$, and HBM-AI communication latency, $L_{HBM-AI}$.


\begin{equation}
    \begin{aligned}
    \label{cyc_per_comm}
        cycle_{comm} = \begin{cases}
            L_{AI-AI} & \text{if data moves from AI to AI}\\
            & \text{chip} \\
            L_{HBM-AI} & \text{if data moves from mem. to} \\
            & \text{AI chip}
        \end{cases}
    \end{aligned}
\end{equation}

\noindent
\textbf{Impact of AI chiplet count.}
As the number of chiplet increases, the physical distance between the source and destination chiplet increases, resulting in increased communication latency. We consider 2D-mesh topology, which is widely used in tile-based architecture for its simplicity and scalability. Routing in the package substrate is more intricate than on-chip. As a result, 
tile-based chiplet architectures have been architected with mesh topology\cite{shao2019simba}. Fig. \ref{fig1:chiplet_alloc_exploration} (b) shows that communication latency increases drastically with the number of chiplets for a mesh topology. 

\begin{figure}[h]
    \centering  
    \includegraphics[scale = 0.775]{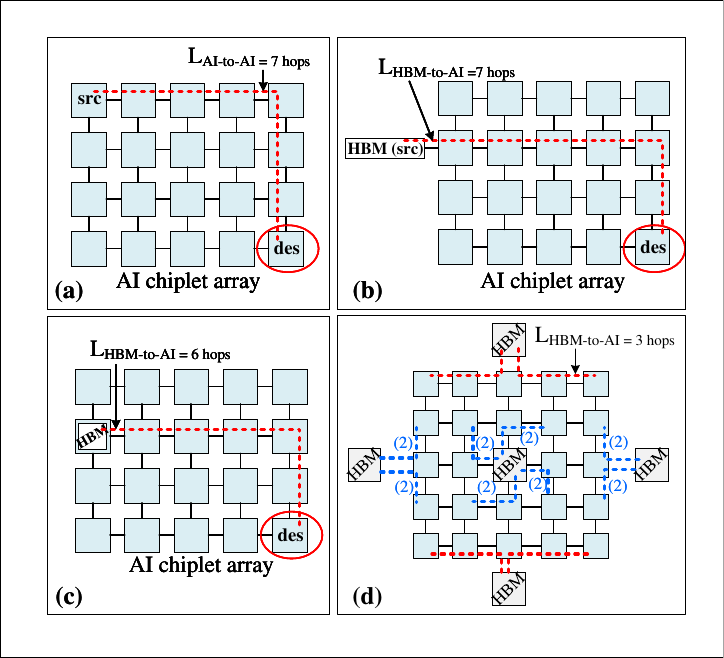}
    \caption{Illustration of latency (in terms of hop) calculation. (a) AI-to-AI chiplet communication, considering the farthest chiplets as source-destination pair. (b) One HBM chiplet, located at the left connected in 2.5D, and the farthest AI chiplet as source-destination pair. (c) One HBM chiplet, 3D-stacked on top of a left-most AI chiplet, and the farthest AI chiplet as source-destination pair. (d) 5 HBM chiplets are placed in 5 different positions. The highest latency decreases from 6 hops (case (c)) to 3 hops with most of the AI chiplets can be provided with data in 2 hops by nearest HBMs.}
    \label{fig:latency}
\end{figure}

\noindent
\textbf{Impact of Chiplet array dimension.}
The longest AI-to-AI chiplet communication latency is expressed as 
\begin{equation}
    \label{eqn:ai2ai_lat}
    \begin{aligned}
        L_{AI-AI} = H_{AI-AI} \times t_{w} + H_{AI-AI} \times  t_{r} + T_{c} + T_{s}
    \end{aligned}
\end{equation}
As we consider a 2D mesh of AI accelerator chiplets, $H_{AI-AI} = m + n -2$, that denotes the number of hops between the source-destination pair.  $m, n$ represents the number of AI chiplets in the X and Y dimension of the array, respectively. $t_{w}$ is per-hop wire delay, $t_{r}, T_{c},$ and $T_{s}$ are router delay, contention delay, and serialization delay, respectively \cite{kite}. Here, $t_{w}, t_{r}, T_{s}$ are design time metrics, that depend on tech. node, interconnect technologies, circuit, and microarchitecture design, $T_{c}$ depends on workload/data traffic, and for a fixed number of chiplets, and routing topology, $H_{AI-AI}$ depends on the chiplet array X and Y dimension.  
We try to keep the aspect ratio of the chiplet array as close as possible to 1 to reduce the communication latency.
In addition, the physical dimension of the chiplet array impacts the system performance by affecting the choice of dataflow and workload mapping strategies \cite{nn_baton}. For a fixed dataflow and mapping strategy, the system performance largely depends on the chiplet array dimension as shown in Fig. \ref{fig:latency}. 

\noindent
\textbf{Impact of HBM\textcolor{black}{/CPU} count and location.}
We analyze the impact of dividing the allocated HBM into multiple chiplets and placing the chiplets in multiple positions on system latency. Partitioning a large chunk of memory into multiple memory chiplets (instead of placing the large memory in one place) and placing these multiple memory chiplets in different locations improves the system latency. Unlike, AI chiplet counts, as the number of HBM chiplets increases, communication latency decreases. Because the communication latency depends on the physical location of the data  \cite{shao2019simba}. Fig. \ref{fig:latency} illustrates how chiplet partitioning and placement improve the system latency. 
As we consider a 2D mesh of AI accelerator chiplets, there are 6 locations: left, right, top, bottom, middle, and 3D stacking, to place the HBM chiplets around the AI chiplets array. These locations result in $2^{6}-1$ combinations for HBM/CPU placements. We model $L_{HBM/CPU-AI}$ same as equation \ref{eqn:ai2ai_lat}, where $H_{AI-AI}$ is replaced by $H_{HBM/CPU-AI}$. 
We use the model presented in \cite{chiplet_glsvlsi} to calculate $H_{HBM/CPU-AI}$ for different locations of \textcolor{black}{HBM/CPU} pair. \textcolor{black}{We consider a 16GB (8-stack, each stack 16Gb) HBM3 chiplet \cite{HBM3}, giving the highest capacity of 80GB with 5 chiplets. We assume that each HBM chiplet has dedicated memory controller integrated within it \cite{AMD_mi300}. As a result, at iso-memory-capacity (i.e, same number of HBMs with integrated memory controller) the cost associated with HBM for both monolithic and chiplet systems is equivalent.}

\textcolor{black}{The host CPU will be primarily responsible for dispatching the workloads to the accelerator chiplets. The package area
will be shared by accelerator chiplet, HBMs as well as CPUs. However, the majority of the package area will be used for AI
computing and HBM memories \cite{AMD_mi300}\cite{shao2019simba}. Hence, in this work we only focus on the AI accelerator and HBMs.}

\begin{figure*}[ht]
    \centering
\includegraphics[width=\textwidth]{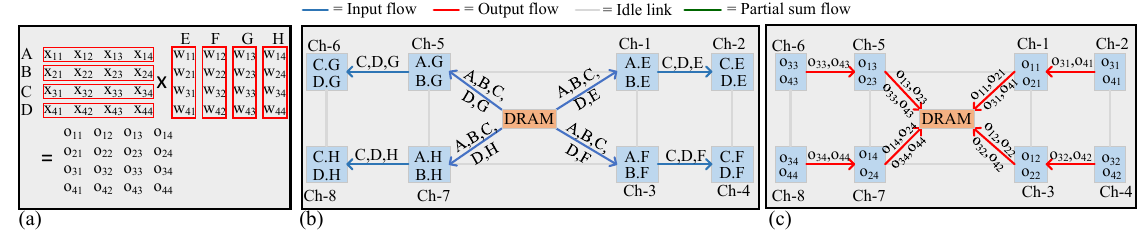}
    \caption{Illustration of mapping and dataflow. (a) Splitting the matrices into smaller parts for different chiplets. (b) Initial data supply from DRAM. Once the chiplets are loaded with required data, computation begins. (c) Final output collection to the DRAM. In this dataflow, there is no inter-chiplet communication during computation for partial sum.}
    \label{fig:dataflow}
\end{figure*}

The above discussion suggests that, for cost-effective integration of more functionalities, we should partition the total chip area into multiple chiplets, each with smaller areas.
From the yield and cost perspective, the more the number of chiplets, the better throughput and less cost. However, this also introduces another consideration: an increase in the number of chiplets results in higher inter-chiplet communication latency, ultimately diminishing throughput and energy efficiency. Therefore, a balance must be struck between dividing the area into an appropriate number of chiplets to enhance functionality and ensure the associated communication latency does not compromise overall system performance and efficiency.

\subsection{Package architectures and configurations}
We explore different packaging architectures, interconnects, and their different configurations 
\cite{ISSCC2021forum}\cite{EMIB}\cite{FOVEROS}\cite{SoIC}
to analyze their impact on the system performance and budget. 


\subsubsection{Inter-chiplet communication bandwidth}
\label{sec:inter_chiplet_comm_bw}
The system utilization term, $U_{sys}$, of equation \ref{eq:ops_sys} depends on the inter-chiplet communication bandwidth. We define $U_{sys}$:
\begin{equation}
    \label{eq:u_sys}
    U_{sys} = \frac{BW_{act}}{BW_{req}}
\end{equation}
Where, $BW_{act}$ is the actual bytes of data transferred per sec and $BW_{req}$ is the required bytes to keep all the neighboring AI chiplets at 100\% utilization, i.e., no stalling for data. For the layout of AI and HBM chiplets we consider in this work, the HBM chiplet needs to deliver data to its 4 neighboring AI chiplets simultaneously at most, and any AI chiplet needs to deliver data to its 1 neighboring chiplets at most. However, it can change with the mapping strategies. \textcolor{black}{As the communication between CPU and AI chiplet primarily involves the instruction dispatch and output accumulation, the communication bandwidth is dominated by bandwidth requirements of the AI accelerator to HBM chiplet.}

\noindent
\textbf{Chiplet mapping exploration.} For large sequence lengths and batch sizes of NLP/LLM models as well for large FC/Conv. layers of DNN models, the matrix sizes get larger, which need to be split temporally in the monolithic chips if the monolithic chip does not contain enough PE units and memory. Having multiple chiplets, the matrices can be split spatially and mapped to multiple chiplets, performing parallel computation.
As illustrated in Fig. \ref{fig:dataflow}(a-c), the input matrix is split along rows (A, B, C, D), and the weight matrix is split along columns (E, F, G, H). Chiplets 1, 3, 5, 7 handle data chunks A and B, while Chiplets 2, 4, 6, 8 handle C and D. The weight matrix portions (E, F, G, H) are distributed to all chiplets accordingly. During initialization, the DRAM supplies data $4\times [A,B,C,D]$, and $[E,F, G, H]$ simultaneously to chiplets 1, 3, 5, 4, with A and B reaching neighboring chiplets in one hop and C and D reaching distant chiplets in the next hop. Data chunks E, F, G, H reach neighboring and distant chiplets in one hop and two hops, respectively. The outputs are collected back to DRAM once the computations are completed. Outputs from neighboring chiplets (ch-1, ch-3, ch-5, ch-7) reach DRAM in one hop, while outputs from distant chiplets reach DRAM in two hops. No inter-chiplet communication is required for partial sum accumulation, however, the required AI-HBM bandwidth (or number of channels) is higher in this mapping strategy, as DRAM needs to broadcast $[A, B, C, D]$ to all four neighboring chiplets. According to the above mentioned mapping and dataflow, the required bandwidth is formulated as 


 \begin{equation}
    \label{eqn:bw_req}
    BW_{req} = 
    \begin{cases}
        4 \times N_{o} \times d_{w}\times f\times (\frac{ops}{sec})_{AI\_chip} & \text{if src. is}\\  
         & \text{HBM} \\
        1 \times N_{o} \times d_{w} \times f\times (\frac{ops}{sec})_{AI\_chip} & \text{if src. is AI} \\
        & \text{chip}  
    \end{cases}
\end{equation}
Where, $N_{o}$ is the number of operands required to perform a MAC operation, which is 2 in general (two multipliers for the multiplication and no new external operands are needed for addition). $d_{w}$ is the data width and $(ops/sec)_{AI\_chip}$ is the peak throughput of the AI chiplet, and $f$ is the frequency of the accelerator. If $BW_{act} \ge BW_{req}$, then there is no stalling in initializing the chiplets' PE array with data. However, if $BW_{act} < BW_{req}$, then there will be $\lceil\frac{BW_{req}}{BW_{act}}\rceil$ cycle stalling for operand data to start the computation. We penalize the overall system throughput with these stalling periods while estimating the system throughput. From equation \ref{eqn:bw_req}, the required bandwidth is smaller if the peak throughput of the AI chiplet is low, resulting in less penalty.


\noindent
\textbf{Impact of Data rates and Link count.}
\noindent
The data rate per pin (in Gbps), $DR$, and the number of links assigned for data transfer, $L$,  of different package type determines the active bandwidth, $BW_{act}$,
\begin{equation}
    \label{eqn:bw_act}
    BW_{act} = DR \times L
\end{equation}
$DR$ and $L$ depend on the interconnect technology. It plays a significant role in the system throughput by contributing to the system utilization. 


\subsubsection{Inter-chiplet communication energy}
As mentioned earlier, interchiplet communication energy, $E_{comm}$ depends on the packaging architecture, we model it as
\begin{equation}
    \begin{aligned}
        \label{energy_comm}
        E_{comm} = E_{bit\_pkg} \times bit_{tot}
    \end{aligned}
\end{equation}
$E_{bit\_pkg}$ is the energy per bit data communication for different interconnect technologies, and $bit_{tot}$ is the data traffic required for the desired operation. 

\noindent
\textbf{Impact of trace length and no. of RDL layers.} 
For a specific data rate and link count, $E_{bit\_pkg}$ again depends on trace length, $tr\_len$, (link-to-link distance between two interconnected dies). To achieve a specified data rate over a longer trace length, intricate circuit techniques and more RDL layers are required resulting in the $E_{bit\_pkg} \propto tr\_len$ relationship \cite{ISSCC2021forum}. 


\subsubsection{Packaging cost}
The packaging cost ($C_{P}$) depends on the packaging architecture and interconnect type. For the same package type, the packaging cost again depends on (i) package area ($A_{P}$), (ii) number of layers (i.e., core and RDL), and (iii) link count (L) and modeled as \cite{package_cost} :

\begin{equation}
    C_{P} = \mu_{0}A_{P} + \mu_{1}L + \mu_{2}
\end{equation}
Where $\mu_{0}, \mu_{1},$ and $\mu_{2}$ are the regression parameters based on the number of core and RD layers. In this work, we consider a fixed package area of 900mm\textsuperscript{2}, leaving the packaging cost dependent on the number of package layers and link density.

The above discussion suggests that, based on the $BW_{req}$, which also depends on the number of chiplets, energy and cost budget,  appropriate allocation of $DR$ and $L$ requires co-optimization, such that the hardware is not suffering from under-utilization while not spending too much budget unnecessarily. 

\section{Optimizing Chiplet-based Architecture}
\label{sec4:optimizer_framework} 
In this section, we build a framework to efficiently navigate the search space, as detailed in Table \ref{tab:design_space}, aiming to optimize throughput, energy, and cost efficiency. \textcolor{black}{Before delving into the details of the proposed optimization methodology and design-space exploration we compare the existing simulators, optimizers, and architectures for AI accelerators from both monolithic and chiplet-based domain in Table \ref{frame_wrk_cmpr}.}

\setlength{\tabcolsep}{2.75pt}
\begin{table}[ht]
\centering
\caption{Parameters and values of Design Space}
\label{tab:design_space}
\begin{tabular}{|c|c|}
\hline
Parameter & Values \\ \hline
Architecture type & 2.5D, 5.5D: (i) memory-on-logic \\
 & (ii) logic-on-logic \\ \hline
No. of chiplets & 1 to 128 @ step of 1 \\ \hline
No. \& location of HBMs & \begin{tabular}[c]{@{}l@{}}Left, right, top, bottom, middle,\\ 3D stacked; 2\textsuperscript{6} -1 location\end{tabular}  \\ \hline
AI2AI interconnect 2.5D & CoWoS, EMIB \\ \hline
AI2AI data rate 2.5D (Gbps) &  1 to 20 @ step of 1 \\ \hline
AI2AI link count 2.5D & 50 to 5000 @ step of 50  \\ \hline
AI2AI trace length (mm) 2.5D & 1 to 10 @ step of 1 \\ \hline
AI2AI interconnect 3D & SoIC, FOVEROS \\ \hline
AI2AI data rate 3D (Gbps)& 20 to 50 @ step of 1 \\ \hline
AI2AI link count 3D & 100 to 10,000 @ step of 100 \\ \hline
AI2HBM interconnect 2.5D &  CoWoS, EMIB \\ \hline
AI2HBM data rate 2.5D (Gbps) &  1 to 20 @ step of 1 \\ \hline
AI2HBM link count 2.5D & 50 to 5000 @ step of 50 \\ \hline
AI2HBM trace length 2.5D (mm) & 1 to 10 @ step of 1 \\ \hline
\end{tabular}
\end{table}

\setlength{\tabcolsep}{2.75pt}
\begin{table*}[ht]
\centering
\tiny
\textcolor{black}{
\caption{Comparative summary of AI accelerator simulator frameworks for design space analysis}
\label{frame_wrk_cmpr}
\begin{tabular}{|l|l|l|l|}
\hline   
\multicolumn{1}{|c|}{}& \begin{tabular}[c]{@{}l@{}}Simulator/Optimizer/\\ Architecture\end{tabular} & \multicolumn{1}{c|}{Design space/Architectural details}& \begin{tabular}[c]{@{}l@{}}Monolithic/\\ Chiplet-based\end{tabular} \\ \hline
Scale-sim \cite{scalesim}& Performance \textbf{Simulator} & \begin{tabular}[c]{@{}l@{}}(i) Dataflow: WS, IS, OS; (ii) HW resource: No. of PE, on-chip memory size; (iii) Architecture type: Eyeriss, TPU\end{tabular} & Monolithic\\ \hline
\begin{tabular}[c]{@{}l@{}}Timeloop+\\ accelergy \cite{timeloop}\end{tabular} & \begin{tabular}[c]{@{}l@{}}Mapping \textbf{optimizer} + Performance, \\ energy, and area \textbf{simulator}\end{tabular} & \begin{tabular}[c]{@{}l@{}} (i) Dataflow: WS, IS, OS; (ii) HW resource: No. of PE, on-chip memory size; (iii) Different memory hierarchies; (iv) Architecture type: Eyeriss, Simba\end{tabular}& Monolithic\\ \hline
Maestro \cite{maestro} & \begin{tabular}[c]{@{}l@{}}Performance and energy \\ \textbf{Simulator} +\textbf{Optimizer}\end{tabular}& \begin{tabular}[c]{@{}l@{}}(i) Dataflow: flexible; (ii) HW resources: Number of PEs, NoC BW/Latency, on-chip memory size; (iii) Architecture type: NVDLA-like\end{tabular}& Monolithic\\ \hline
Astra-sim \cite{astrasim}& Performance \textbf{Simulator}& \begin{tabular}[c]{@{}l@{}} (i) Distributed training: data, model, hybrid parallelism; (ii) Hierarchical collective algorithms: on-load, off-load; (iii) Fabric design: number of links \\ \& latency/BW per link; (iv) Fabric topology: pt-to-pt, 2D/3D Torus\end{tabular} & \begin{tabular}[c]{@{}l@{}}Multi-chip \\ (package/board level)\end{tabular} \\ \hline
STONNE \cite{stonne} & \begin{tabular}[c]{@{}l@{}}Performance, energy and area \\ \textbf{Simulator}\end{tabular}                        & \begin{tabular}[c]{@{}l@{}}Architecture type: flexible \& reconfigurable architecture\end{tabular}& Monolithic \\ \hline
Confucuix\cite{kao2020confuciux} & \begin{tabular}[c]{@{}l@{}}Performance and energy \\ \textbf{optimizer}\end{tabular}& (i) HW resource: Number of PE, on-chip buffer size& Monolithic\\ \hline
SIMBA \cite{shao2019simba}& \begin{tabular}[c]{@{}l@{}}Performance \& power \\ \textbf{Optimizer} + 36-chiplet \textbf{Architecture}\end{tabular}    & \begin{tabular}[c]{@{}l@{}} (i) Mapping and tiling strategies; (ii) 36 NVDLA chiplets connected in 2D mesh\end{tabular}& \begin{tabular}[c]{@{}l@{}}Multi-chip \\ (package level)\end{tabular}\\ \hline
SPRINT \cite{sprint}& 64-chiplet \textbf{Architecure}& \begin{tabular}[c]{@{}l@{}}64-chiplet architecture with photonic interconnect for interchiplet communication\end{tabular}& \begin{tabular}[c]{@{}l@{}}Multi-chip \\ (package level)\end{tabular}       \\ \hline
NN-Baton \cite{nn_baton}& \textbf{Simulator} + \textbf{Optimizer}& \begin{tabular}[c]{@{}l@{}}(i) No. of accelerator chiplets: 1, 2, 4, 8; (ii) On-chip memory: 36 to 642KB; (iii) Different mapping strategies; (iv) Routing topology: Ring\end{tabular}& \begin{tabular}[c]{@{}l@{}}Multi-chip\\ (package level)\end{tabular}\\ \hline
Moand\cite{monad} & \textbf{Optimizer} & \begin{tabular}[c]{@{}l@{}}(i) No. of accelerator chiplets; (ii) mapping \& tiling; (iii) packaging; (iv) network topology; (v) placement\end{tabular}& \begin{tabular}[c]{@{}l@{}}Multi-chip\\ (package level)\end{tabular}       \\ \hline
TVLSI'20\cite{arch_chip_pckg_codsgn_flw} & \textbf{Simulator}& \begin{tabular}[c]{@{}l@{}} (i) 64-core ROCKET-64 architecture; (ii) 2.5D interposer-based centralized NoC \end{tabular}& \begin{tabular}[c]{@{}l@{}}Multi-chip\\ (package level)\end{tabular}        \\ \hline
\begin{tabular}[c]{@{}l@{}}Chiplet-gym \\ (This work) \end{tabular}& \begin{tabular}[c]{@{}l@{}}Performance, power, area and \\ cost \textbf{Optimizer}\end{tabular}                   & \begin{tabular}[c]{@{}l@{}} (i) No. of AI accelerator chiplets, no. \& location of HBM chiplets, (ii) package architecture: 2.5D, 3D; (iii) Interconnect types \& configuration \\(details in Table \ref{tab:design_space})\end{tabular}& \begin{tabular}[c]{@{}l@{}}Multi-chip \\ (package level)\end{tabular}       \\ \hline
\end{tabular}}
\end{table*}

Comprising of 14 parameters and their possible values, our parameter space has more than $2\times 10^{17}$ design points which poses challenges for exhaustive search due to its time and resource-intensive nature. To address this, we explore learning-based and meta-heuristic search approaches to efficiently reach global or near-global optima.

\begin{figure*}[ht]
    \centering
    \includegraphics[width=0.95\textwidth, height=7.2cm]{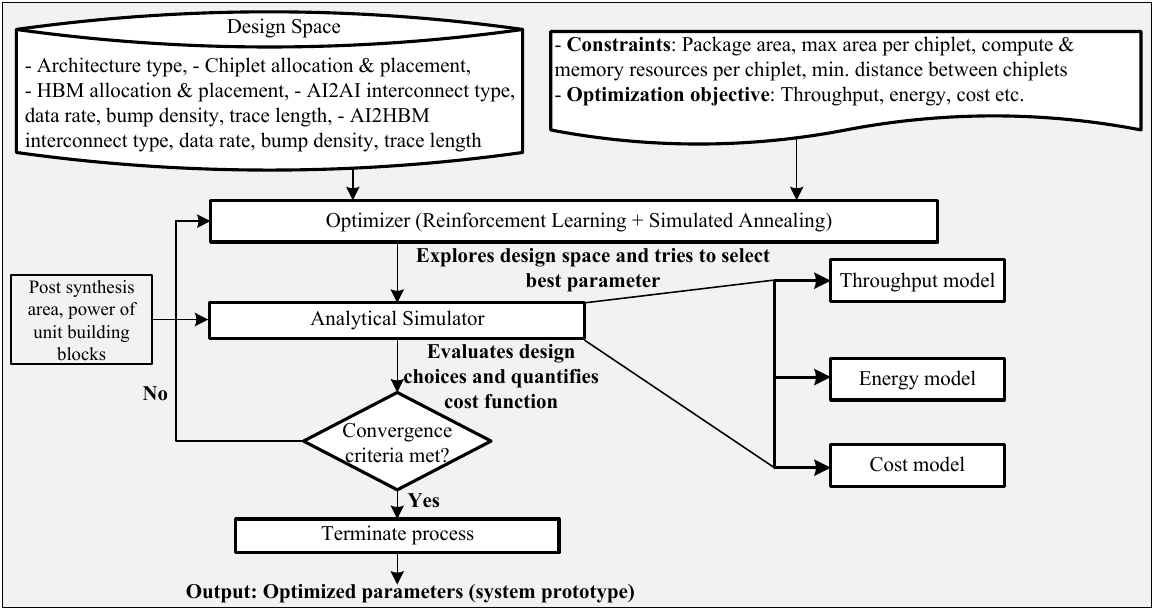}
    \caption{Optimization framework overview}
    \label{fig:opt_framework}
\end{figure*}

Because of the inherent stochastic nature of Reinforcement Learning (RL) and Simulated Annealing (SA) algorithms, we observe slight variations in the achieved objective function values. To enhance the robustness of the optimizer, we train multiple RL models and SA algorithms with different seed values. Subsequently, we perform an exhaustive search across the outcomes of these algorithms to pinpoint the optimum solution (refer to Alg. \ref{alg:alg_top}). An overview of the optimization framework is presented in Fig. \ref{fig:opt_framework}. It takes the design space and constraints as input and outputs the optimized design points. 

\RestyleAlgo{ruled}
\SetKwComment{Comment}{/* }{ */}
\begin{algorithm}[ht]
    \caption{Proposed optimization algorithm}
    \label{alg:alg_top}
$t \gets Trial_{max}$\;
$obj_{best} \gets -inf$\;
\While{$t \leq Trial_{max}$}{
$param_{SA}, obj_{SA} \gets SA()$\;
\If{$obj_{SA} > obj_{best}$}{
$param_{best}, obj_{best} \gets param_{SA}, obj_{SA}$\;
}
$param_{RL}, obj_{RL} \gets RL()$\;
\If{$obj_{RL} > obj_{best}$}{
$param_{best}, obj_{best} \gets param_{RL}, obj_{RL}$\;
}
    }
\Return $param_{best}, obj_{best}$
\end{algorithm}

\subsection{RL problem formulation}
RL tries to mimic human learning behavior to learn about a new environment. In RL, an \textbf{Agent} continuously interacts with an \textbf{Environment}, takes \textbf{Actions} by observing the present \textbf{State} of the environment, receives feedback as a form of \textbf{Reward} from the environment, and updates its underlying \textbf{Policy} to take new actions to maximize reward. After enough interaction with the environment, the agent can take a specific set (sequence) of actions that maximize the reward in the given environment. 
Formulating a Markov Decision Process (MDP) consisting of a tuple of five key elements: <$\mathcal{S}, \mathcal{A},\mathcal{P},r, \mathbb{\gamma}>$ is at the core of formulating an RL problem. Where $\mathcal{S}$ = State space, $\mathcal{A}$ =  Action space, $\mathcal{P}$ = Transition probability matrix of going to $S_{t}$ from $S_{t-1}$ by taking action $A_{t-1}$, $r$ = Reward function, and $\gamma$ = discount factor that takes any value between $[0,1]$ \cite{sutton2018reinforcement}.

\noindent
\textbf{Environment} provides feedback to the agent by quantifying the rewards. In our case, we incorporate our analytical expressions discussed in Section \ref{sec3:thruput_energy_formulation} into a Gym\cite{openai_gym} environment, known as Chiplet-Gym, to assess the performance of the action taken by the agent. 

\noindent
\textbf{State or Observation space} contains the set of all possible states of the environment. It should include all the information for the agent to take the next action, making the process an MDP. In our case, the observation space contains the following items: \{\emph{maximum package area, the maximum area allowed per chiplet, current area per chiplet, ai2ai communication latency, ai2hbm communication latency, current communication energy, current packaging cost, current throughput}\}.

\noindent
\textbf{Action space} defines the set of all possible actions available to the agent each time step. Our action space, consisting of a combination of discrete integers and categorical values, corresponds to the parameters we aim to optimize. Given the state of the environment and the reward, the agent selects values for each of the parameters in Table \ref{tab:design_space} to maximize the reward.


\noindent
\textbf{Reward} is provided to the agent as a form of feedback in response to every action it takes. We formulate the reward function same as the objective function we want to maximize
\begin{equation}
\label{eq:final_cost_func}
    r = \alpha T - \beta C - \gamma E
\end{equation}
Where $T, C, E$ represent the throughput, packaging cost, and communication energy, respectively. $\alpha, \beta, \gamma$ are the user-defined constants that let the users put specific weight on specific parameters of the objections function, such as throughput, cost, energy-efficiency during optimization. \textcolor{black}{Based on the reward, which is formulated from the analytical expressions of Section \ref{sec3:thruput_energy_formulation}, RL finds the optimum design choices considering complex trade-offs of chiplet area, bandwidth, chiplet-to-chiplet communication.}

\noindent
\textbf{RL algorithm} We use Proximal Policy Optimization (PPO) algorithm \cite{ppo} implemented by Stable-Baselines3 \cite{stbl_baseline} because of its simplicity, computational efficiency, and compatibility with the action and state space of our problem. PPO is a on-policy policy gradient method that combines the idea of having multiple workers from Advantage Actor-Critic (A2C) algorithm and the idea of using trust region to improve the current policy from Trust Region Policy Optimization (TRPO) algorithm  \cite{stbl_baseline}.

\subsection{Simulated Annealing}
In addition to RL, we also explore meta-heuristic search approaches, such as simulated annealing,  to evaluate their efficacy in navigating the design space. Simulated annealing adds an exploitation step on top of random search. It randomly samples the design points and in addition to accepting the better design points, based on the acceptance criterion, it also accepts the design points that worsen the objective function. We modify the simulated annealing algorithm by slightly changing the acceptance criterion for our problem. The algorithm is shown in Algorithm \ref{alg:sim_anling}. We use the same objective function of Equation \ref{eq:final_cost_func}  to optimize.


\RestyleAlgo{ruled}
\SetKwComment{Comment}{/* }{ */}
\begin{algorithm}[ht]
    \caption{Modified simulated annealing algorithm}
    \label{alg:sim_anling}
$iteration \gets T_{max}$\;
$temp \gets temperature$\;
$st\_sz \gets step\_size$\;
$X_{curr} \gets$ randomly choose initial solution\;
$O_{curr} \gets$  evaluate initial solution\;
$X_{best}, O_{best} \gets  X_{curr}, O_{best}$\;
\While{$iterations \leq T_{max}$}{
\tcc{find candidate solution}
    $X_{cand} \gets X_{curr} + uniform(-1,1)*st\_sz$\;
\tcc{evaluate candidate solution}
    $O_{cand} \gets f(X_{cand})$\;
    \If{$O_{cand}>O_{best}$}{
    $O_{best} \gets O_{cand}$\;
    $X_{best} \gets X_{best}$\;
    }
    $t \gets temp/iterations$\;
    \If{$O_{cand}> O_{curr}$ \textbf{OR} $rand()<t $}{
    $X_{curr}, O_{curr} \gets X_{cand}, O_{cand}$\;
    }
    }
\Return $X_{best},O_{best}$
\end{algorithm}

Finally, we deploy RL and SA algorithm multiple times, followed by conducting a comprehensive search on the outputs produced by SA and RL agents.

\textcolor{black}{While demonstrated explicitly for AI accelerators and mesh routing topology, the proposed optimization framework can be generalized to diverse chiplet-based designs and routing topology, requiring users to model their architectures and network topology in equation 4, 10, 11, and 13, to find the correct blend of package and interconnect architecture. For example, I/O chiplets provide signal transmission and regeneration. Their performance can be modeled as extra latency in our framework.}

\section{Experiments and Results}
\label{sec5:experiment_and_result}
In this section we discuss experimental method, implementation details, and results. 
\subsection{Experimental method}
\label{subsec:exp_method}
As shown in Fig. \ref{fig:opt_framework}, at the core of the optimizer we implement PPO and simulated annealing algorithm. The optimizer explores the design space and tries to select the best parameters sticking to the design constraints and user-given optimization objective, such as throughput optimization, energy, and/or cost optimization. To evaluate the optimizer's objective function, we implement our cost model, explained in Section \ref{sec3:thruput_energy_formulation}, in an OpenAI Gym\cite{openai_gym} environment named as Chiplet-Gym.


\begin{table}[ht]
    \centering
    \caption{Per hop wire length and delay for 2.5D and 3D architecture \cite{wire_len_delay}\cite{EMIB}} 
    \begin{tabular}{|c|c|c|} \hline
        Packaging arch. & Per hop wire length (mm) & Delay, $t_{w}$ (ps) \\ \hline
        2.5D & 1 & 17.2 \\ \hline
        3D & 0.08 & 1.6 \\ \hline
    \end{tabular}
    
    \label{tab:wire_len_delay}
\end{table}

\begin{table}[ht]
\centering
\caption{Interconnects' properties\cite{ISSCC2021forum}}
 \setlength{\tabcolsep}{2.25pt}
\begin{tabular}{|c|c|c|c|c|}
\hline
Interconnect & Bond/bump & TSV pitch & Energy & Implementation \\ 
 & pitch ($\mu m$)& pitch($\mu m$) & (pJ/bit) & cost \\ \hline
CoWoS & 30 - 40  & - & 0.2 $\sim$0.5 &  Medium  \\ \hline
EMIB & 55 - 45  & - & 0.17 $\sim$0.7  & Low \\ \hline
SoIC & 9  & 9 & 0.1 $\sim$0.2  & High \\ \hline
FOVEROS & $<10 $ & - & $<0.05$ & Highest \\ \hline
\end{tabular}

\label{tab:interconnect_properties}
\end{table}
We consider a fixed amount of package area, 900mm\textsuperscript{2}, dedicated for AI and HBM chiplets \cite{AMD_mi300}. To avoid thermal hotspot, we place the chiplets at 1mm apart from each other in a mesh topology \cite{chiplet_spacing}. This leaves ($900-(m+n+2)mm^{2}$) of area for the chiplets. 
The optimizer will select the number of chiplets such that it maximizes the throughput while sticking to the area constraint. The area per chiplet is calculated as the total package area available for AI chiplets over the number of chiplets. Analyzing the yield vs area curve (Fig. \ref{fig1:chiplet_alloc_exploration}) we set the maximum allowable area per chiplet to 400mm\textsuperscript{2} as a constraint. Because, at 14nm, for the die area beyond 400mm\textsuperscript{2} the yield is even lower than 75\%. Inspired by the recent trend of higher on-chip memory to reduce the DRAM accesses \cite{tpuv4}, we allocate 40\% of the chiplet area to the compute resources, 40\% to the on-chip SRAM, and rest 20\% to other blocks such as control, IO, NoC, routing etc. For 3D architecture, we have to sacrifice some of the area of the chiplet for the TSV and its associated keepout zone. From SoIC TSV pitch of 9um\cite{ISSCC2021forum}, >12K TSVs can be fit into 1mm\textsuperscript{2}. So we keep at most 2mm\textsuperscript{2} for TSV in 3D architecture. Which is enough for both signal and power supply \cite{intact}. We use the values shown in Table \ref{tab:wire_len_delay} and \ref{tab:interconnect_properties} in our throughput, cost, and energy model to calculate the cost function of the design points. 


\subsection{Implementation details}
The optimization framework\footnote{https://github.com/KFM135/chiplet-optimizer} is written in Python v3.9. and run on an Intel hexa-core i5-9500 @ 3 GHz machine.

\subsubsection{RL}
The Chiplet-Gym environment is constructed by integrating our analytical simulator into OpenAI Gym v0.26.2 \cite{openai_gym} to establish a unified interface between the RL algorithm and the analytical simulator. We define the action space as MultiDiscrete and observation space as Box space. The simulator receives the RL policy's action (i.e., a combination of various parameters forming a design point) as input and produces corresponding throughput, energy, and cost values. The environment's state is then updated, and the reward is calculated. The state and reward are fed to the agent, enabling it to adjust its network to maximize rewards for subsequent actions. 

\noindent
\textbf{Policy-Value network.} PPO utilizes the Multi-Layer Perceptron (MLP) as both of its policy and value network. The architecture of the actor or policy network is defined as [10, 64, 64, 810], and the architecture of the critic or value network is set as [10, 64, 64, 1], employing the tanh activation function. The size of the input for both networks is determined by the dimension of the observation space, while the output layer size of the policy network is determined by the action space. The output layer size of the value network is set to 1.

\begin{figure}[ht]
    \centering
    \includegraphics[width=0.48\textwidth]{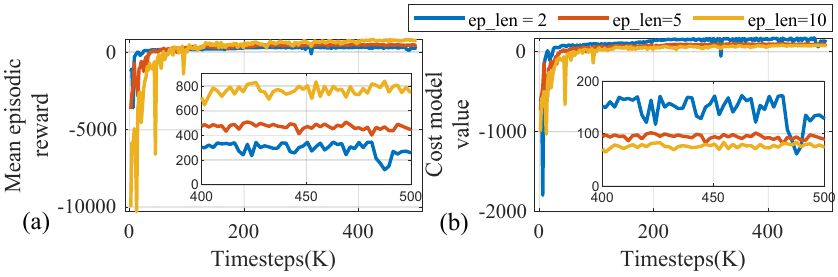}
    \caption{Impact of episode length in convergence (PPO algorithm). Inset shows the zoomed-in version of each plot.}
    \label{fig:impact_of_ep_len}
\end{figure}

\noindent   
\textbf{Impact of episode length on RL convergence.} The algorithms are trained with an episode length of 2. While a longer episode length often results in a higher mean episodic reward, it does not guarantee a superior cost model value for the optimized parameters. Although longer episodes are generally associated with increased exploration, our hypothesis is that, in our specific case, the agents lean towards exploitation to maximize rewards. This hypothesis arises from the fact that our reward values span from a large negative value to a positive one. Once the agent discovers a positive value, it tends to exploit that particular action to maximize the mean episodic reward neglecting further exploration of the design space. Figure \ref{fig:impact_of_ep_len} (a)
shows that the agent achieves a mean episodic reward of 800 at episode length of 10, where as the cost model value of these actions are less than 100 (Fig. \ref{fig:impact_of_ep_len} (b)). On contrary, at episode length of 2, the mean episodic reward is around 300 and the cost model value is around 150. (Note: The cost model value at each timesteps are calculated as $mean\_episodic\_reward/episode\_length$.)

\begin{figure}[ht]
    \centering
    \includegraphics[width=0.48\textwidth]{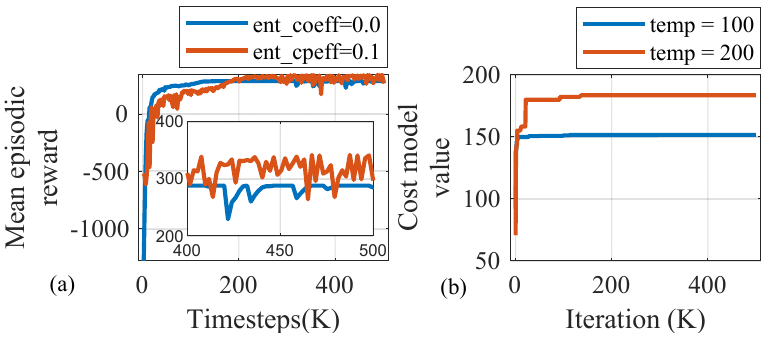}
    \caption{(a) Impact of entropy coefficient in RL convergence and (b) impact of temperatue on SA convergence . Inset shows the zoomed-in version of each plot.}
    \label{fig:impact_of_ent_coeff}
\end{figure}

\noindent
\textbf{Impact of entropy coefficient on RL convergence} Another hyper-parameter impacting the exploration and exploitation balance is entropy coefficient. Serving as a regularizer, entropy coefficient plays a crucial role in shaping the behavior of the RL agent during training. A larger entropy coefficient implies that all actions are equally likely, fostering exploration, while a smaller entropy coefficient indicates that one action's probability within the policy dominates, emphasizing exploitation.
Fig. \ref{fig:impact_of_ent_coeff} (a) shows that when entropy coefficient is set to 0, the agent stabilizes to a lower reward value more rapidly. However, when the entropy coefficient is increased to 0.1, the agent achieves a higher reward value, albeit with a slightly less stable trajectory. In this case, we use an entropy coefficient of 0.1 to reach higher convergence value. Other 
significant hyperparameters of PPO algorithm are shown in Table \ref{tab:ppo_hyper_param}.



\begin{table}[ht]
\centering
    \caption{PPO hyper-parameters \& their values}
\begin{tabular}{|c|c|c|c|}
\hline
n\_steps & 2048& n\_epoch & 10 \\ \hline
batch\_size & 64 & learning rate & 0.0003 \\ \hline
clip range & 0.2 & value func. coef. & 0.5  \\ \hline
entropy\_coeff. & 0.1 & discount factor & 0.99 \\ \hline
\multicolumn{3}{|c|}{bias-variance trade-off factor} & 0.95 \\ \hline
\end{tabular}
\label{tab:ppo_hyper_param}
\end{table}

\subsubsection{Simulated Annealing}

We employ Algorithm \ref{alg:sim_anling}, initializing it with a randomly chosen candidate solution from the design space. Like PPO, SA's performance is also sensitive to initial temperature, a measure of exploration vs exploitation. As shown is Fig. \ref{fig:impact_of_ent_coeff} (b), SA achieves significant higher cost model value with higher temperature value. Higher temperature value ensures more exploration by increasing the probability of accepting a worse trial point. As a result, the initial temperature to 200, and a step size of 10 is employed for locating the neighboring points. We do not use the general Metropolis acceptance criterion, $metropolis = exp - \{(O_{curr} - O_{cand})/t\}$, due to the potential for $(O_{curr} - O_{cand})$ to become very large or very small, leading to the $metropolis$ evaluating to either infinity or 0. Instead, we solely utilize the parameter $t$ to statistically accept poorer solutions in the early stages, facilitating exploration of the search space. $O_{curr}$=cost model value for current design point and $O_{cand}$=cost model value at candidate design point.

    


\subsection{Results}
\subsubsection{Performance and runtime analysis of optimizer}

\begin{table*}[ht]
\centering
\caption{Optimized parameters for $\alpha,\beta,\gamma=[1,1,0.1]$ found by PPO algorithm}
\begin{tabular}{|c|c|c|}
\hline
Parameter & Case (i): 64 chiplets as upper bound & Case (ii): 128 chiplets as upper bound \\ \hline
Architecture type & 5.5D-Logic-on-Logic & 5.5D-Logic-on-Logic \\ \hline
No. of chiplets & 60 (30 3D chiplet pairs arranged in 5X6 2.5D mesh) & 112 (56 3D chiplet pairs arranged in 7X8 2.5D mesh \\ \hline
No. \& location of HBMs & \begin{tabular}[c]{@{}c@{}}4 HBM chiplets @ top, bottom, right, and \\ middle of 5X6 chiplet pairs\end{tabular} & \begin{tabular}[c]{@{}c@{}}4 HBM chiplets @ left, right, bottom, and \\ middle of 7X8 chiplet pairs\end{tabular} \\ \hline
AI2AI interconnect 2.5D & EMIB & EMIB \\ \hline
AI2AI data rate 2.5D & 20 Gbps & 20 Gbps \\ \hline
AI2AI link density 2.5D & 3100 & 1450 \\ \hline
AI2AI trace length 2.5D & 1 mm & 1 mm \\ \hline
AI2AI interconnect 3D & SoIC & FOVEROS \\ \hline
AI2AI data rate 3D & 42 Gbps & 34 Gbps \\ \hline
AI2AI link density 3D & 3200 & 4400 \\ \hline
AI2HBM interconnect 2.5D & EMIB & EMIB \\ \hline
AI2HBM data rate 2.5D & 20 Gbps & 20 Gbps \\ \hline
AI2HBM link density 2.5D & 4900 & 3850 \\ \hline
AI2HBM trace length 2.5D & 1 mm & 1 mm \\ \hline
\end{tabular}
\label{tab:opt_param}
\end{table*}

\begin{figure}
    \centering
    \includegraphics[width=0.48\textwidth]{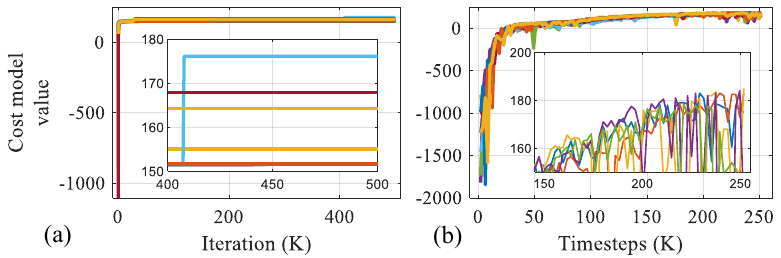}
    \caption{Convergence behavior of (a) SA and (b) RL for multiple runs with \textcolor{black}{10} different seed values for case (i) (i.e., 64 chiplets). Inset shows the zoomed-in version of each plot.}
    \label{fig:SA_RL_64_chiplet_multirun}
\end{figure}

\begin{figure}
    \centering
    \includegraphics[width=0.48\textwidth]{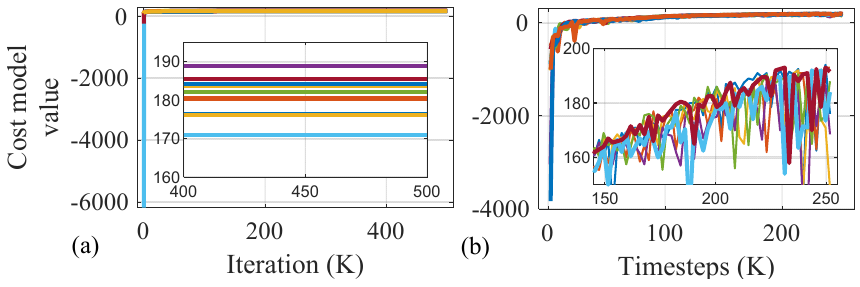}
    \caption{Convergence behavior of (a) SA and (b) RL for multiple runs with \textcolor{black}{10} different seed values for case (ii) (i.e., 128 chiplets).  Inset shows the zoomed-in version of each plot.}
    \label{fig:SA_RL_128_chiplet_multirun}
\end{figure}

\begin{figure}
    \centering
    \includegraphics[width=0.48\textwidth]{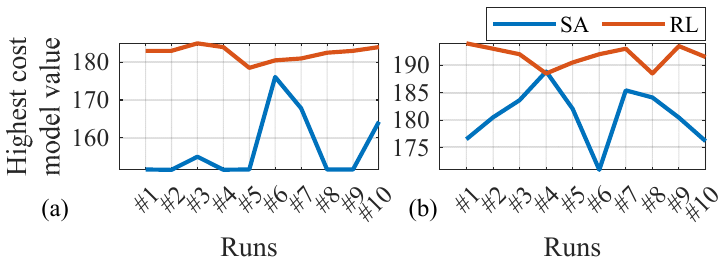}
    \caption{Highest cost model value achieved by the SA and RL algorithms for multiple runs: (a) for 64 chiplets and (b) for 128 chiplets.}
\label{fig:SA_RL_highest_cost_multiple_run}
\end{figure}
In our investigation of the design space, we consider two distinct scenarios: case (i), wherein the upper limit for the number of AI chiplets is set to 64, and case (ii), where this upper limit is increased to 128. We ran each of the algorithms multiple times for each cases with different seed values to ensure their convergence stability. Fig. \ref{fig:SA_RL_64_chiplet_multirun} and \ref{fig:SA_RL_128_chiplet_multirun} (a) and (b) show the convergence behavior of SA and PPO algorithm for case (i) and case (ii) for \textcolor{black}{10 runs}, respectively. As expected, both algorithms achieve a better cost model value for case (ii) because of its higher throughput, however, due to large packaging cost, case (i) 64 chiplets as the upper bound, is considered more practical. Fig. \ref{fig:SA_RL_highest_cost_multiple_run} (a) and (b) show the highest cost model value achieved by SA and RL algorithm over 10 runs for case (i) and (ii), respectively. We observe that RL achieves higher cost model values each run and more stable over multiple runs ranging from 178 - 185 for case (i) and 188 - 194 for case (ii). Where SA achieves 151 - 176 and 170 - 188 for case (i) and case (ii), respectively.  

The run time of SA for 500K iterations is less than a minute and the run time to train the PPO agent for 250K timesteps is <20 mins. We finally integrated several trained RL agents and performed SA optimization on-the-go, and performed an exhaustive search among those SA and RL agents. The final optimizer with 20 SAs and 20 RL trained RL agents take around 10 mins to report the optimized parameter. As the RL is used in inference mode, here the SA dominates the runtime.    


\subsubsection{Optimized architecture evaluation}


\begin{figure*}
    \centering
    \includegraphics[width=\textwidth]{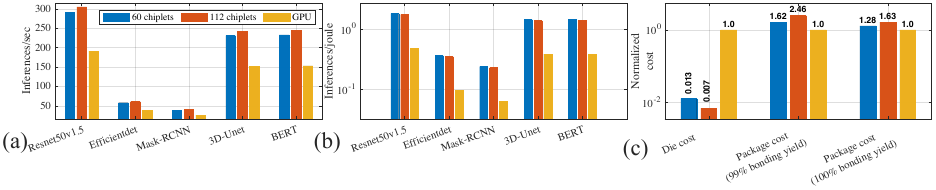}
    \caption{Comparison between 60-chiplet, 128-chiplet, and monolithic system. (a) Inferences/sec, (b) Inferences/joule for MLPerf benchmark, and (c) cost.}
    \label{fig:chiplet_vs_gpu_all}
\end{figure*}



Table \ref{tab:opt_param} shows the optimized parameter found by the optimizer for both cases for a specific $\alpha, \beta, \gamma$ value (\textcolor{black}{user-defined weights on the objective function as explained in Eqn. \ref{eq:final_cost_func}}). We observe that best parameter was found by RL PPO algorithm. It is imperative to note that multiple design configurations may exist, achieving almost identical (less than 1 variation) cost model value.

The optimum design point for case (i) consists of thirty 3D AI chiplet pairs arranged in a $5\times 6$ mesh topology, resulting in 60 chiplets in total. 2 chiplets (forming a pair) are connected with SoIC 3D integration technology with a data rate of 42Gbps per link and link count of 3200 providing up to 131.25 Tbps of bandwidth. Each chiplet pair is connected with other chiplet pair with 2.5D EMIB integration with a data rate of 20Gbps and a link count of 3100 delivering up to 60 Tbps of bandwidth. Four 16GB HBM chiplets, located at top, right, bottom, and middle of the $5\times 6$ mesh topology, are connected to two to four neighboring AI chiplets with EMIB 2.5D integration technology with a data rate of 20Gbps per link and a link count of 4900, resulting in a bandwidth of 95 Tbps. The trace length for each 2.5D interconnect is selected as the minimum trace length possible (minimum chiplet-to-chiplet distance). In case (ii), when we increase the maximum number of chiplets to 128, we observe that the optimum design configuration contains 112 chiplets (56 chiplet pairs) and the communication bandwidth decreases for all cases. This is because, as the number of chiplets increases, area per chiplet decreases, resulting in smaller throughput per chiplet, less bandwidth demand, and high system utilization. We observe that 3D architecture, even with area penalty for TSV and TSV-associated keep-out zone\cite{3d_gtech}, achieves $1.52\times$ more logic density than its 2D/2.5D counterpart at the same package size.

We synthesize the chiplet module, found by the optimizer, with Synopsys Fusion Compiler using their 14nm PDK \cite{saed14} at 1GHz clock frequency and obtain the peak throughput per chiplet, $(ops/sec)_{AI\_chip}$, and energy consumption per MAC operation, $E_{op*}$. We use these values in our analytical model to estimate the throughput and energy efficiency of the 60 and 112 chiplet system. For cost estimation we use the model from \cite{chiplet_actuary}.

\begin{table}[ht]
\tiny
\centering
\caption{DNN benchmark features}\textcolor{black}{
\begin{tabular}{|c|c|c|c|}
\hline
\begin{tabular}[c]{@{}c@{}}Benchmark model\end{tabular}& Domain& Dataset        & \begin{tabular}[c]{@{}c@{}}Operations/forward pass task\end{tabular} \\ \hline
Resnet50& \begin{tabular}[c]{@{}c@{}}Image classification\end{tabular}& Imagenet& 4 GFLOPs\\ \hline
Efficientdet& \begin{tabular}[c]{@{}c@{}}Light weight object detection\end{tabular} & COCO 2017& 410 GFLOPs\\ \hline
mask-RCNN& Heavy weight object detection& COCO 2014      & 447 GFLOPs\\ \hline
3D-UNet& Biomedical image segmentation& KiTS19         & 947 GFLOPs\\ \hline
\begin{tabular}[c]{@{}c@{}}BERT \end{tabular} & Natural Language Processing& Wikipedia 2020 & 32 GFLOPs\\ \hline
\end{tabular}}
\label{tbl:bench_mark_features}
\end{table}

Fig. \ref{fig:chiplet_vs_gpu_all} compares the 60-chiplet, 112-chiplet and monolithic GPU for MLPerf benchmark \cite{mlperf}. 
Chiplet-based 3D system achieves higher throughput for all the benchmarks than monolithic chip because of its $1.52\times$ more logic density (Fig. \ref{fig:chiplet_vs_gpu_all} (a)). \textcolor{black}{The benchmark features are briefly summarized in Table \ref{tbl:bench_mark_features}.}
The 112-chiplet system experiences higher communication latency compared to the 60-chiplet system. However, the lower bandwidth penalty of the 112-chiplet system, in contrast to the 60-chiplet system, outweighs the higher latency, resulting in a superior overall throughput compared to the 60-chiplet system. 60 and 112-chiplet systems are $3.7\times$ and $3.6\times$ more energy-efficient (inverse of energy consumption) than the monolithic counterpart, respectively (Fig. \ref{fig:chiplet_vs_gpu_all}(b)). 
\textcolor{black}{The fact that chiplet architectures, which requires comparatively slower inter-chiplet links provide 1.5x more
throughput, and 3.7x energy efficiency compared to the monolithic one, is counter-intuitive. Because, there is no better solution than housing as many logic and cache as possible on the same die, given the hypothetical scenario that there is no yield, cost, or reticle size limitations (however this is not practically achievable). On the contrary, due to 3D stacking, the logic density, hence the throughput, has increased at the same area footprint. To achieve the same amount of throughput with the monolithic chip, more than one monolithic chip need to be linked off-board on the PCB, which consumes at least one order of magnitude more energy\cite{off_board_power} than on-package communication. This is the reason, 3D-stacked chiplet based architecture offers 1.5x throughput and 3.7x energy-efficiency
compared to its monolithic counterpart.}

\textcolor{black}{Fig. \ref{fig:chiplet_vs_gpu_all}(c) illustrates that the die cost of the monolithic chip is $76\times$ and $143\times$ higher than that of the 60 and 112 chiplet systems, respectively. This significant cost difference arises from the low yield (48\%) of the monolithic chip of 826mm\textsuperscript{2}, compared to the 97\% and 98\% die yield of the 60 and 112 chiplet systems, with a die size of 26mm\textsuperscript{2} and 14mm\textsuperscript{2}, respectively, at 7nm node. In addition to that, the cost of the cost of Known Good Dies (KGD) is inversely proportional to the number of KGD ($N_{KGD}$). As the die area (A) increases, the number of good dies ($N_{KGD}$) decreases, leading to a substantial increase in cost. The relationship between the cost and die area can be approximated as $cost_{KGD} \propto A^{\frac{5}{2}}$ (taking up to 2 terms of Taylor series expansion of die yield) \cite{off_board_power}\cite{chiplet_actuary}. Please note that, we implement our chiplet in synopsys 14nm free PDK. However, use estimate the cost for 7nm to have a fair comparison between the monolithic one, which was fabricated in 7nm technode\cite{amp100_gpu}.}

\textcolor{black}{We consider chiplet I/O pad bonding yield of $99\%$ to estimate the packaging cost. In fact with better process control this bonding yield can reach $100\%$ as reported by TSMC in  \cite{SoIC}. Furthermore, even if there are faulty bondings,  $100\%$ TSV/Hybrid-Bonding bonding yield can be achieved by using the TSV/pad repair techniques with spare TSV/pad \cite{tsv_repair}. We observe that the packaging cost of chiplet-based architecture is higher, $1.62\times$ and $2.46\times$, than the monolithic one. However, with near perfect TSV bonding yield\cite{SoIC}\cite{tsv_repair}, the packaging cost of chiplet-based architecture is only $1.28\times$ (for 60-chiplets) and $1.63\times$ of the monolithic one. Despite the package cost of the chiplet-based architecture being $1.62\times$ ($1.28\times$ for $100\%$ bonding yield) higher than that of the monolithic architecture,
the advantages of smaller dies in the chiplet-based design offset the additional package expenses, as the silicon die cost contributes most to the overall manufacturing RE cost.}

\section{Related works}
\label{sec6:related_works}
\subsection{Chiplet-based architecture exploration}
\subsubsection{DNN accelerator}
SIMBA \cite{shao2019simba} is a pioneering work in chiplet-based AI accelerator, that integrates 36 NVDLA-like accelerator chiplets on a package. Centaur \cite{centaur} integrates CPU and FPGA chiplets on package, specially for recommendation system workload. SPRINT \cite{sprint} is a 64-chiplet system with photonic interconnect for DNN inference. There have been few works in chiplet based architecture focusing on different aspect of design space exploration.
NN-Baton\cite{nn_baton} proposes a framework for DNN workload mapping and chiplet granularity in small scale (1 to 8 chiplets), however, they do not consider the packaging integration aspect and fabrication cost. While Monad \cite{monad} incorporates mapping, resource allocation, communication and different package substrate to optimize for PPA and fabrication cost, their packaging integration design space is limited to 2.5D, excluding 3D. \cite{chiplet_cloud} proposes ChipletCloud for LLM inference, however, their chiplets are connected in board-level instead of package level. 

\subsubsection{General purpose} Some works focus on the exploration of Network-on-Package (NoP) and reliable routing protocols \cite{NoP_chiplet}
for chiplet-based architecture. \cite{2.5D_placement}
explore network topology and cost-aware chiplet placement for 2.5D architecture. \cite{chiplet_actuary} puts forward a cost model for evaluating the 2.5D manufacturing cost. \cite{off_board_power}\cite{package_reticle_limit} suggest the importance of chiplet design space exploration for performance, energy, cost, reliability enhancement. 

\subsection{RL in Design-space exploration}
Deep Reinforcement Learning has gained popularity in exploring the design space exploration and optimization of the EDA domain, spanning from front-end (i.e., planning and architectural exploration) \cite{usenix}\cite{RL_NoC}\cite{MARL_DRAM}\cite{kao2020confuciux}
 to back-end (i.e., implementation, physical design and circuit design)\cite{googlechipRL}\cite{gcnRL} \cite{autocktRL}. To the best of our knowledge, this work is the first to perform a comprehensive design space search, encompassing resource allocation, placement, packaging architectures (both 2.5D and 3D), and their configurations to optimize for Power, Performance, Area, and Cost (PPAC) using Deep Reinforcement Learning (DRL).
\textcolor{black}
{\section{Limitations and Future Works}
\label{sec7:lim_n_future_wrks}
\textcolor{black}{Even though our parameter space contains more than $2\times 10^{17}$ design points, it does not capture all possible scenarios. In order to keep the design space concise and tractable, we limit our design space as mentioned in Table \ref{tab:design_space} and made several assumptions as mentioned in Section \ref{subsec:exp_method}. Exploring other routing topology such as p2p with photonic interconnects, H tree, bus, ring etc., exploring more heterogeneous architectures, multi-tier 3D-stacks, placement of host CPU chiplets and exploring their different layouts are future works.}}
\section{Conclusion}
\label{sec8:conclusion}
This paper proposes Chiplet-Gym to explore the design space of chiplet-based AI accelerators to optimize for Power, Performance, Area, and Cost (PPAC). To evaluate the design points, we analytically model the power, performance, and cost for chiplet-based AI accelerator. With reinforcement learning and simulated annealing, the optimizer is robust and efficient in locating the global or near-global optima of the design space for PPAC. Results show that the optimizer finds the design point that achieves 1.52$\times$ throughput, 0.27$\times$ energy, and $0.01\times$ die cost while incurring only $1.62\times$ package cost of its monolithic counterpart at iso-area.


%





\ifCLASSOPTIONcaptionsoff
  \newpage
\fi



\bibliographystyle{IEEEtran}
\bibliography{main_ref}

\begin{IEEEbiography}[{\includegraphics[width=1\linewidth]{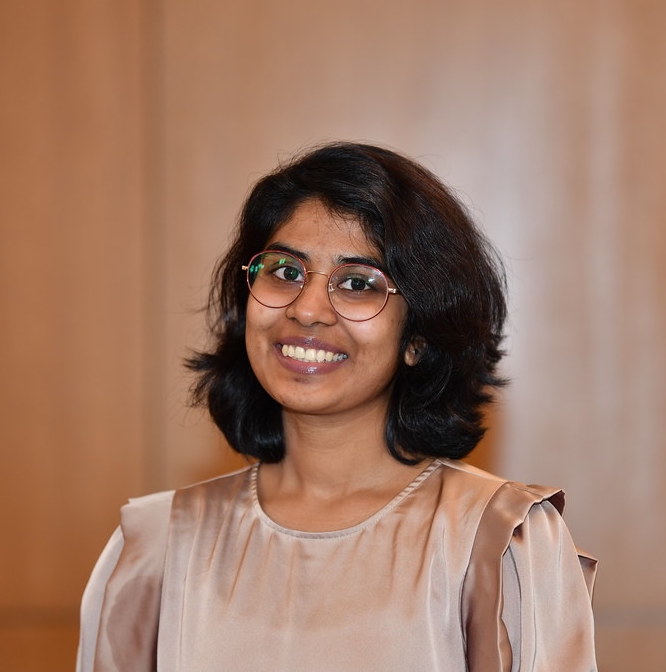}}]{Kaniz Mishty} received the B.S. degree in Electronics and Communication Engineering from Khulna University of Engineering and Technology, Bangladesh, in 2018. She is currently working towards her Ph.D. degree in ECE at Auburn University, AL, USA. Her research interests are energy efficient AI hardware design and AI/ML in CAD. She interned with Apple Inc. in Summer '22, 23 and Qualcomm in '21   on AI application in SoC and custom circuit design.
\end{IEEEbiography}

\begin{IEEEbiography}[{\includegraphics[width=\linewidth]{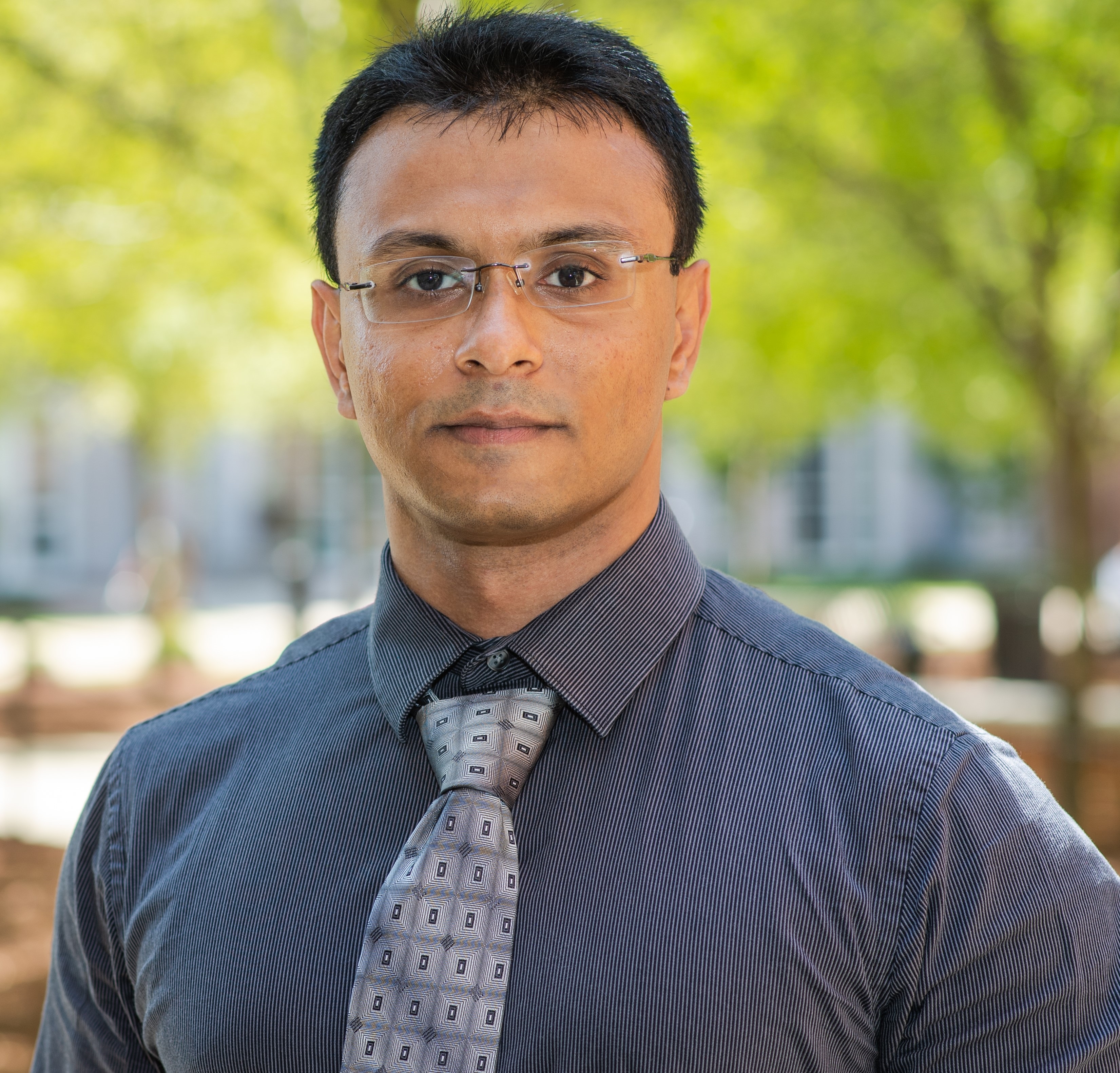}}]{Mehdi Sadi} (S'12-M'17) is an Assistant Professor at the Department of Electrical and Computer Engineering (ECE) at Auburn University, Auburn, AL.  Dr. Sadi  earned his PhD in ECE from  University of Florida, Gainesville, USA in 2017, MS from University of California at Riverside, USA in 2011 and BS from Bangladesh University of Engineering and Technology in 2010. Prior to joining Auburn University, he was a Senior R\&D SoC Design Engineer in the Xeon server CPU design team at Intel Corporation in Oregon. Dr. Sadi`s research focus is on developing algorithms and CAD techniques for implementation, design, reliability, and security of AI, and brain-inspired computing hardware. His research also spans into developing Machine Learning/AI enabled design flows for  System-on-Chip (SoC), and Design-for-Robustness for safety-critical AI hardware systems. He has published more than 25 peer-reviewed research papers. He was the recipient of Semiconductor Research Corporation best in session award and Intel Xeon Design Group recognition awards.
\end{IEEEbiography}








\end{document}